\documentclass[a4,twocolumn]{revtex4}
\usepackage{graphicx}
\usepackage{epstopdf}
\usepackage{longtable}

\begin{document}

\title{Solving the Einstein-Podolsky-Rosen puzzle: the origin of non-locality in Aspect-type experiments.}

\author{Werner A. Hofer\\
        Department of Physics, University of Liverpool\\
L69 3BX Liverpool, United Kingdom}

\begin{abstract}
So far no mechanism is known, which could connect the two measurements in an Aspect-type experiment. Here, we suggest such a mechanism, based on the phase of a photon's field during propagation. We show that two polarization measurements are correlated, even if no signal passes from one point of measurement to the other. The non-local connection of a photon pair is the result of its origin at a common source, where the two fields acquire a well defined phase difference. Therefore, it is not actually a non-local effect in any conventional sense. We expect that the
model and the detailed analysis it allows will have a major impact on quantum cryptography and quantum computation.
\end{abstract} %\pacs{03.65.Ud}

keywords: entanglement, Bell inequalities, coincidence measurements, Einstein-Podolsky-Rosen paradox

\maketitle

{\it Introduction} -- One of the most puzzling results in modern physics, based originally on a paper by Einstein, Podolsky, and Rosen (EPR)\cite{EPR}, is the apparent non-locality of correlation measurements in quantum optics \cite{aspect99}. The experiments performed on pairs
of entangled photons, beginning with the experiments by Alain Aspect in 1982 \cite{aspect82}, seem to prove beyond doubt that the two measurements are not independent. The measurements are usually interpreted in terms of the Bell inequalities \cite{bell64}, which assert that their violation, corresponding to the experimental results and also the theoretical predictions of quantum optics, amounts to a non-local connection between the two independent measurements \cite{aspect99}.
There has been much debate on whether such a non-local connection implies superluminal effects, see for example
Maudlin's book \cite{maudlin94}. The present consensus is that in these experiments no information travels faster than light
from one point of measurement to the other.

Here, we want to approach the subject from a new angle. Rather than
analyzing the problem, whether quantum optics is complete, which was the focus of the original EPR paper \cite{EPR},
we start from the assumption that quantum optics contains, in its mathematical formalism, the answer to the problem. In consequence, the conceptual difficulty so far has been due to the lack of transparency in its mathematical framework. We
can formulate this hypothesis in two distinct statements:
\begin{enumerate}
\item Quantum optics is complete.
\item The connection between the two photons is due to their common source.
\end{enumerate}
The second statement relates to the fact that a common source for both photons of an entangled pair is
a common feature in {\it all} Aspect-type experiments. These pairs always have a common origin at
one and only one source. It will be seen that these simple insights are sufficient to develop
a model of photon entanglement accounting for all features in the experiments.

{\it Theoretical model} -- One could formulate the problem either in terms of a photon's spin, or in terms of its electromagnetic fields. In our model we choose the electromagnetic field as the fundamental property. That photons,
or light waves with limited extensions, do possess electromagnetic fields is also undoubtedly true.
Light waves and electromagnetic fields imply periodic amplitudes of a photon's fields, they also imply a
wavelength $\lambda$ and a phase $\phi$ during photon propagation. That phases also play a role in the experiments
is witnessed by the fact that the two photons need to be coherent, i.e. they need to have a coherent
phase relation for correlation effects to be observable. This was one of the main limitations in
early experiments, since optical fibres then were not of sufficiently high quality to preclude thermal
effects and a randomization of phases over relatively short distances. Experiments with time-like
separation thus were only possible about twenty years after the first experiments by Aspect.

\begin{figure}
\centering
\includegraphics[width=\columnwidth]{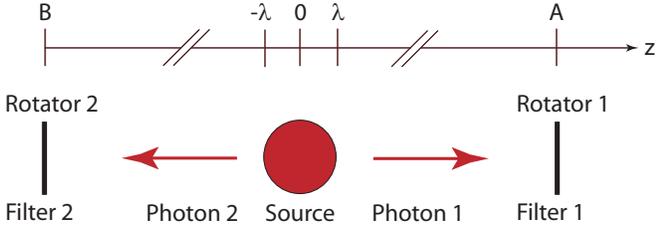}
\caption{{\bf Aspect-type experiment.} Two photons are emitted from a common source, and subjected at two points $A$ and $B$ to separate polarization measurements, here a rotation of the fields and a filter.}\label{Fig1}
\end{figure}

At this point we have electromagnetic fields traveling from a common source to their points of measurement, while
retaining a coherent phase relative to their point of origin (see Figure \ref{Fig1}). Our next problem is the
development of a mathematical model for the polarization measurements.
In this respect, it is well known that the Pauli matrices $\hat{\sigma}_i$ are a matrix description of rotations in three
dimensional space, which in Clifford Algebra or Geometric Algebra are described by geometric products.
This is due to their common algebra \cite{lasenby02}:
\begin{eqnarray}
{\bf e}_i {\bf e}_j = \delta_{ij} + i \epsilon_{ijk} {\bf e}_k \nonumber \\
\hat{\sigma}_i \hat{\sigma} _j = \delta_{ij} + i \epsilon_{ijk} \hat{\sigma}_k
\end{eqnarray}
Here, the ${\bf e}_i$ are framevectors of three dimensional space, and $\epsilon_{ijk}$ is the
Levi-Civita permutation tensor.
It is also known that polarization measurements in optics correspond to a rotation of the
electromagnetic fields, since the angle of rotation, which is proportional to a potential in
a non-linear crystal, is one of the input parameters of the experiments. It is therefore quite
natural to model a polarization measurement by a rotation in geometric algebra, with a rotation
axis parallel to the direction of motion. As this creates complex numbers, if the rotation is
modeled as a geometric product acting on an exponent, it is then required to use a filter in
the model which limits the measured results to real values. In this way all components
of a Bell-type experiment can be modeled locally, and it can be analyzed, from this simple and
transparent mathematical model, where the observed non-locality actually comes from.

In Figure \ref{Fig1} we show the setup of a Aspect-type experiment. Two photons with a defined phase difference $\Delta$ are emitted from a common source. After traveling an arbitrary number of wavelengths $\lambda$ of their associated field, either to $A$, in positive $z$-direction, or to $B$, in negative $z$-direction, they are subjected to a measurement, which is assumed to consist of a rotation of the photon fields and a filtering. Rotations in geometric algebra \cite{lasenby02} are described by a multiplication with a geometric product of two vectors. Here, we assume rotations perpendicular to the direction of photon propagation, which act on a Poynting-like vector of the electromagnetic fields. The rotations are then described by:
\begin{eqnarray}
R(A) &=& R\left(z_1\right) = \exp \left({\bf e}_1{\bf e}_2\right) {\bf e}_3 z_1\,2 \pi/\lambda \nonumber \\
R(B) &=& R\left(z_2\right) = \exp -\left({\bf e}_1{\bf e}_2\right) {\bf e}_3 z_2\,2 \pi/\lambda
\end{eqnarray}
where the values of $z_i$ are limited by $0 \le z_i \le \lambda$. The rotations thus cover all values from zero to a full
rotation of 2$\pi$. It is evident that the rotations are local measurements, i.e. the rotation at point $A$ is independent of the rotation at point $B$. Given that the geometric products involve a product of the three frame vectors ${\bf e}_i$, the brackets can be omitted and the triple product $\left({\bf e}_1{\bf e}_2\right) {\bf e}_3 = {\bf e}_1{\bf e}_2 {\bf e}_3 = i$. The two rotations are thus:
\begin{eqnarray}
R\left(\varphi_1\right) &=& \exp i z_1\,2 \pi/\lambda = e^{i\varphi_1} \nonumber \\
R\left(\varphi_2\right) &=& \exp -i \left(z_2\,2 \pi/\lambda + \Delta\right) = e^{-i\left(\varphi_2 + \Delta\right)}
\end{eqnarray}
where we symbolized the product $z_i 2\pi/\lambda$ by $\varphi_i$. The normalized probability $p$ of detecting photons after
a rotation with angle $\varphi_i$ shall be given by the square of the real part of the rotation, or:
\begin{equation}
p(\varphi_i) = \left[\Re (R(\varphi_i))\right]^2
\end{equation}
The probability in this case models a filter, acting after the rotator. Here, the measurement depends on the phase difference
between the source of the photon pair and the end point of the rotation. The real part of the phase difference is thus the square root of the detection probability. For coincidence measurements we therefore have to consider the phase difference between the two end points of the rotation of both photons. The correlations between two measurements with angles $\varphi_1,\varphi_2$ are then described by a square of the real part of the product:
\begin{equation}
p\left(\varphi_1,\varphi_2\right) = \left[\Re \left[R(\varphi_1) \cdot R(\varphi_2)\right]\right]^2
\end{equation}
The relations between rotations and probabilities of photon measurements are:
\begin{eqnarray}
p(\varphi_1) &=& \cos^2 \varphi_1 \qquad p(\varphi_2) = \cos^2 \varphi_2 \nonumber \\
p(\varphi_1,\varphi_2) &=& \cos^2 (\varphi_1 - \varphi_2 - \Delta)
\end{eqnarray}

The framework can be generalized to three and more rotations. Assuming that we have two rotators on either side,
positioned at arbitrary locations along the photon's paths, the conditional probability for four individual measurements
with rotators  $\varphi_1$ to $\varphi_4$, where $\varphi_1$ and $\varphi_2$  are in positive $z$-direction while
$\varphi_3$ and $\varphi_4$ are in negative $z$-direction, is equal to:
\begin{equation}
p(\varphi_1,\varphi_2,\varphi_3,\varphi_4) = \cos^2 (\varphi_1 + \varphi_2 - \varphi_3 - \varphi_4 - \Delta)
\end{equation}
These predictions, which go well beyond current models, can easily be checked by future experiments on entangled photons. It should also be noted that the complex phase is not limited to rotators in the measurements, but that a complex phase
connects the two photons in every case where the field vectors rotate. This occurs in practically all experiments
undertaken so far.

From a statistical point of view an initial phase $\varphi_0$ at the source of both photons does alter the outcome of single
polarization measurements, but does not affect correlations. To appreciate this aspect of the model, let us assume that the pair of photons possesses an initial phase at the source, described by $S_{0,i}$, which is unknown and statistically distributed
within the interval [$0, 2\pi$]:
\begin{equation}
p(S_{0,i}) = constant = \frac{1}{N} \qquad  0 < S_{0,i} < 2 \pi
\end{equation}
where $N$ is the number of coincidence measurements taken. Then the probability to measure a photon at, say, $A$ will also be statistically distributed with the probability given by:
\begin{equation}
p_i(\varphi_{1}) = \cos^2 \left(\varphi_{1} + S_{0,i}\right) \qquad i = 1,N
\end{equation}
The same statistical behavior will be observed at $B$, since:
\begin{equation}
p_i(\varphi_{2}) = \cos^2 \left(\varphi_{2} + S_{0,i}\right) \qquad i = 1,N
\end{equation}
However, the correlations will not be affected and remain independent of this initial phase, since:
\begin{eqnarray}
p_i(\varphi_1,\varphi_2) &=& \cos^2 (\varphi_1 + S_{0,i} - \varphi_2 - S_{0,i} - \Delta) \nonumber \\
&=& \cos^2 (\varphi_1 - \varphi_2 - \Delta)
\end{eqnarray}

{\it Correlations} -- To appreciate the novelty of the approach it is illuminating to cite Alain Aspect's review paper in 1999
\cite{aspect99}: ''The violation of Bell's inequality, with strict relativistic separation between the chosen
measurements, means that it is impossible to maintain the image '\'a la Einstein' where correlations are explained
by common properties determined at the common source and subsequently carried along by each photon. We must
conclude that an entangled EPR photon pair is a non-separable object; that is, it is impossible to assign
individual local properties (local physical reality) to each photon. In some sense, both photons keep in
contact through space and time.''

Here, we found that the ''common property ... carried along by each photon'' is a complex phase, which
will be altered in a polarizer. The actual normalized count $p$ does not reveal the full physical situation;
it is therefore necessary to take the correlated normalized count for the product of two complex rotations
and not, as assumed in the derivation of Bell's inequalities, the product of the two separate normalized
counts. The additional information about the imaginary component of the phase is not revealed in the local
counts, even though it is present in the local rotations. It seems thus that the difference between the
physical process involved (a rotation of the fields and a filtering), and the actual measurement result
(a count of photons after
rotation), has not been appreciated to date. Rotations in three dimensional space, formalized within the
framework of geometric algebra, are the key to understanding spin properties of electrons \cite{lasenby02,hofer11}.
Based on this analysis, it seems that they are equally key to understanding
polarizations of photons and electromagnetic fields.

{\it Experiments} -- Experimentally, the measurements are performed on a pair of down-converted photons \cite{aspect82}, which are separated, subjected to a polarizer,
and then measured either in a spin-up or spin-down state at the detectors. Experiments are usually interpreted
in terms of the Clauser-Horne-Shimony-Holt-inequalities (CHSH)\cite{CHSH}, which are based on normalized expectation
values $E(\varphi_1,\varphi_2)$, derived from coincidence counts of photon spins at the two points of measurement.
Within the present context it is actually unnecessary to define exactly, what spin-up and spin-down means in a
measurement; it suffices to assume that they will be subject to the same relation between rotational angles
and detection probability. For a phase-difference $\Delta = 0$ the normalized detection rates for spin-up and
spin-down photons will be (we denote coincidences by a capital C, as is standard in the literature, and also use
the convention that a coincidence is the measurement of equal spin for both, spin-up (+) and spin-down (-) components):
\begin{eqnarray}
C^{++} &=& C^{--} = \cos^2\left(\varphi_1 - \varphi_2\right) \nonumber \\
C^{+-} &=& C^{-+} = 1 - \cos^2\left(\varphi_1 - \varphi_2\right)
\end{eqnarray}
Then we obtain the same correlations of polarizations as in Aspect's first experiments \cite{aspect82}, namely:
\begin{equation}
E(\varphi_1,\varphi_2) = 2 \cos^2\left(\varphi_1 - \varphi_2\right) - 1 = \cos 2\left(\varphi_1 - \varphi_2\right)
\end{equation}
Correlations at different pairs of angles $\varphi_1,\varphi_2$ can be combined to a sum $S$, which, according to
Bell's derivation \cite{bell64}, should not be larger than two for any local model. Within the present model we
obtain, in accordance with experimental results and also with predictions of quantum optics:
\begin{eqnarray}
S\left(\varphi_1,\varphi_1',\varphi_2,\varphi_2'\right) &=& E(\varphi_1,\varphi_2) - E(\varphi_1,\varphi_2') + \\ &+&
E(\varphi_1',\varphi_2) + E(\varphi_1',\varphi_2') = 2 \sqrt{2} \nonumber
\end{eqnarray}
if  $\varphi_1 = 0, \varphi_1' = 45, \varphi_2  = 22.5, \varphi_2' = 67.5$, in violation of the Bell inequalities.
The model thus fully accounts for experimental values under ideal conditions (which are nearly reached in the most
advanced experiments \cite{weihs99}), and also for the standard predictions in quantum optics.

{\it Origin of nonlocality} -- The underlying reason that quantum optics appears to be non-local is its formulation in terms of operators and expectation values which entail integrations over the whole system. A local model, based on geometric algebra and
phases, can obtain the same numerical results, as shown here concerning Aspect-type experiments.
Moreover, while the standard model makes the actual connection of entangled
photons somewhat less than transparent, the model developed has the advantage that all processes are local and transparent.
There is, as shown, no connection between the two measurements exceeding the speed of light. In addition, the present
model is also a model of photon entanglement. All that is required for entanglement, it seems, is a coherent
phase between the two photons. Whether this model is the whole answer to the problem, or only part of an eventually fully
comprehensive theory, cannot be estimated at present. This will to a large extent depend on subsequent experimental tests.

 In this context it is interesting to note that the present analysis is based only on the field properties of
photons and their rotational features. It does not need to consider any particle properties to arrive at the derived results.
However, it does also not specify, whether an individual photon at a particular setting of the polarizer will actually be
measured or not. For the macroscopic outcome such a specification is neither necessary, nor does it form part of any
theoretical model which describes the experiments at present. It is certainly not part of quantum optics, which, as
shown, can be replicated with a model based on three-dimensional rotations and phases of the photon fields. But it is
also not contained in Bell's analysis of the original EPR problem \cite{bell64}, where it is never specified, what will trigger the
detection of individual photons. Eventually, this question might be answered by a detailed analysis of the dynamical
processes in the polarizer (rotator/filter in the present model) itself, which contains hidden variables in the exact
shape of the fields or the thermal fluctuations of the polarizer atoms. Such a detailed picture is not necessary, though,
to establish the correlations which have been puzzling physicists for more than thirty years.

{\it EPR paradox} -- Returning to the original EPR problem and the question whether there is an ''element of reality'' in the experiments, which is not described by the formalism in quantum optics, it turns out that Einstein was wrong: the description
in quantum optics via Pauli matrices is an equivalent way of accounting for rotations in three dimensional space.
 The formalism in quantum optics is thus complete. However, Einstein was also right, because the
imaginary component of the phase difference, which is a consequence of rotations in geometric algebra, is not strictly
speaking a physical property of the system in quantum optics, and it is not revealed in the experiments, where only
the real component shows up in the photon count. It is thus a hidden variable. But this imaginary component is a - classical -
geometric component of rotations in geometric algebra, thus it does have physical reality. This reality has so far been
denied in quantum optics. In this sense one could say that even though the framework is formally complete, its relation
to physical reality has not been correctly described. This seems to account also for Bell's derivation of his famous
inequalities \cite{bell64}: since Bell did not ascribe physical reality to the imaginary component of the phase, he
did arrive only at a limited description of the situation from the viewpoint of geometric algebra. Thus his inequalities
can be violated in a local and realistic model, even though it has frequently been asserted that this is not possible.

{\it Quantum steering} -- Even though the model developed in this paper is statistical, it allows, under certain conditions, for the prediction of the experimental outcome at $B$, if the result at $A$ is known. It thus sheds also some light on the mysterious property of ''quantum steering'' \cite{wittmann12}, which  according to Schr\"odinger accounts for the fact that entanglement would seem to allow an experimenter to remotely ''steer'' the state of a distant system as in the Einstein-Podolsky-Rosen (EPR) argument. It will turn out, that this interpretation of experimental results contains in reality
a well known logical fallacy (cum hoc ergo propter hoc). 

To understand in detail, how this is possible, let us assume that the difference between $\varphi_1$ and $\varphi_2$ is 45 degrees: $\varphi_1 - \varphi_2 = \pi/4$. The correlation probability then is $p(\varphi_1,\varphi_2) = 0.5$. Assuming now that $A$ measures a spin-up photon (+), then the result at $B$ is with 50\% probability spin-up (+), and with 50\% spin-down (-). In this case, a prediction of $B$ given $A$ is impossible. The same is true for a difference in angles of 135, 225, 315 degree. In all these cases an experimental outcome at $A$ cannot ''steer'' the outcome at $B$. However, the situation changes drastically if the difference is a multiple of $\pi/2$. In this case the probabilities will be:
\begin{equation}
p(\varphi_1 - \varphi_2 = \pi/2) = 0 \qquad p(\varphi_1 - \varphi_2 = \pi) = 1
\end{equation}
Consider now a change of angle $\varphi_1$ during the experimental run, so that in one case, $\varphi_{1,0}$ the difference between the two angles is $\pi/2$, in the other case, $\varphi_{1,1}$ the difference is $\pi$. It is still impossible to predict the result at $A$, since this will depend, as the single polarization measurement, on the initial phase. However, it is possible to predict the outcome at $B$ if $A$ is known. The measured results at $A$ and $B$ will be:
\begin{eqnarray}
\begin{array}{cccc}
\varphi_{1,0}: &  A = + & \Rightarrow & B = - \\
               &  A = - & \Rightarrow & B = + \\
\varphi_{1,1}: &  A = + & \Rightarrow & B = + \\
               &  A = - & \Rightarrow & B = -
\end{array}
\end{eqnarray}
Given that the experimental outcome of $B$ depends on both, the setting of angle $\varphi_1$ and the outcome at $A$, it is
claimed, in conventional accounts of the experimental data \cite{wittmann12}, that the experiments are the consequence of
some {\it spooky action at a distance} \cite{EPR}. Since the angle can be changed in-flight, also the experimental result of $B$ given a result $A$ seems to change in this time-span. However, the result at $A$ is not known initially, neither is the result at $B$. Therefore the change of the angle $\varphi_1$ does not influence the outcome at $B$, it only affects the correlation. As this analysis shows, the effect is neither: neither spooky, nor action at a distance, but the result of a logical fallacy, called {\it cum hoc ergo propter hoc} in the classics.

We described the two limiting cases, either a certain prediction of outcomes at $B$ given $A$ or a random result $B$ given a particular outcome at $A$. The ability to predict $B$ clearly depends on the angles in the polarization measurements. For varying differences between the two angles we thus get a variable degree of certainty in our predictions. This result of the present model, which is also well beyond current concepts, can easily be checked by experiments.

{\it Outlook} -- Photon entanglement is crucially important for a number of fields like quantum teleportation, quantum cryptography, and quantum computation. In all these fields we expect that this result, and the clarification it presents with respect to the actual mechanism of entanglement and non-locality, will lead to a much better understanding of scientific issues.

{\bf Acknowledgements} The author acknowledges support from the Royal Society London.

\end{document}